\documentclass[structabstract]{aa}
\usepackage{graphicx}
\usepackage{txfonts}
\usepackage{longtable}

\begin{document}
%
   \title{Planetary companions orbiting M giants HD~208527 and HD~220074\thanks{Based on observations made with the BOES instrument on the 1.8-m telescope at Bohyunsan Optical Astronomy Observatory in Korea.}}
   \author{B.-C. Lee\inst{1},
          I. Han\inst{1},
          \and
          M.-G. Park\inst{2}
          }

   \institute{Korea Astronomy and Space Science Institute, 776,
		Daedeokdae-Ro, Youseong-Gu, Daejeon 305-348, Korea\\
	      \email{[bclee;iwhan]@kasi.re.kr}
	    \and
	      Department of Astronomy and Atmospheric Sciences,
	      Kyungpook National University, Daegu 702-701, Korea\\
	      \email{mgp@knu.ac.kr}
             }

   \date{Received 29 Auguest 2012 / Accepted xx xxx 2012}


  \abstract
   {}
   {The purpose of the present study is to research the origin of planetary companions by using a precise radial velocity (RV) survey.
   }
   {The high-resolution spectroscopy of the fiber-fed Bohyunsan Observatory Echelle Spectrograph (BOES) at Bohyunsan Optical Astronomy Observatory (BOAO) is used from September 2008 to June 2012.
   }
   {We report the detection of two exoplanets in orbit around HD 208527 and HD 220074 exhibiting periodic variations in RV of 875.5 $\pm$ 5.8 and 672.1 $\pm$ 3.7 days. The RV variations are not apparently related to the surface inhomogeneities and a Keplerian motion of the planetary companion is the most likely explanation. Assuming possible stellar masses of 1.6 $\pm$ 0.4 and 1.2 $\pm$ 0.3 $M_{\odot}$, we obtain the minimum masses for the exoplanets of 9.9 $\pm$ 1.7 and 11.1 $\pm$ 1.8 $\it M_\mathrm{Jup}$ around HD 208527 and HD 220074 with an orbital semi-major axis of 2.1 $\pm$ 0.2 and 1.6 $\pm$ 0.1 AU and an eccentricity of 0.08 and 0.14, respectively. We also find that the previously known spectral classification of HD 208527 and HD 220074 was in error: Our new estimation of stellar parameters suggest that both HD 208527 and HD 220074 are M giants. Therefore, HD 208527 and HD 220074 are so far the first candidate M giants to harbor a planetary companion.
   }
   {}


   \authorrunning{B.-C. Lee et al.}
   \titlerunning{Planetary companions orbiting M giants HD208527 and HD 220074}
   \maketitle
%

\section{Introduction}

Most exoplanets have so far been detected by using spectroscopic radial velocity (RV) and photometric transit methods at the ratio of two to one. While the transit method is limited by using strongly biased towards short-period companions with a period as short as a few days, the RV technique is relatively less restricted, it can detect planetary systems with a period approximately from a day to $\sim$ 38 years (i.e. 47 Uma described in Gregory \& Fischer 2010).
한글

It has been demonstrated that the RV method has a limited range of stellar spectral types of stars chosen as targets. Planet searches generally focus on late F, G, and K dwarfs because they are bright enough to obtain a high signal-to-noise ratio (S/N) in the high-resolution spectra and have an ample number of spectral lines for the RV measurements. Furthermore, finding planets around the Sun-like stars is a big issue in every aspect. Unfortunately, there are no more candidate stars in the main-sequence (MS) stage for ground-based observations. Recognizing the importance of exoplanet formation and its evolution, several exoplanet survey groups have begun exploring evolved G, K giants (Frink et al. 2002; Setiawan et al. 2003; Sato et al. 2003; Hatzes et al. 2005; D{\"o}linger et al. 2007; Han et al. 2010) and low-mass (M dwarf) stars (Endl et al. 2003; Bonfils et al. 2011).

A giant star has sharper spectral lines than a dwarf star, which allow more precise measurement of RV shift. RV surveys of giant stars are also interesting because we can detect planets of stars that are considerably more massive than the Sun. For MS stars of the same mass this would be difficult because they rotate fast. Compared to MS stars, however, giants show blended RV variations due to various surface variations in the line profiles, such as chromospheric activities, and stellar pulsations. It thus makes it difficult for the giant star to purify the RV origin through the examination process. Also, it is not so easy to determine the masses of the host stars, because stars of different mass are located in roughly the same region of the H-R Diagram. And, the interpretation of the data is not easy, because only massive planets with relatively long orbital periods have been detected for giant stars, whereas surveys of MS stars often find low-mass planets with short orbital periods.

Of the exoplanets detected by the RV method, just over 12\% are discovered around giants. However, no planet has been discovered around M giant stars because they are rare or maybe because planet formation is weak around M giants unlike M dwarfs, which may have a significant probability (up to 13\%) of harboring short-period giant planets (Endl et al. 2003). Short-term oscillations on M giants are rarely observed, yet they show periods in excess of one day (Koen \& Laney 2000). Three-dimensional simulations (Freytag \& H{\"o}fner 2008) predict typical photometric periods of about several tens to a few hundred days, similar to those of RV variations in K giants.

%
\begin{table*}
\begin{center}
\caption[]{Stellar parameters for the stars analyzed in the present paper.}
\label{tab1}
\begin{tabular}{lccc}
\hline
\hline
    Parameters              & HD 208527     & HD 220074 &    Reference     \\

\hline
    Spectral type             & K5 V    & K1 V        & Hipparcos (ESA 1997)   \\
                                   & --      & M2 III      & Kidger et al. (2003); Tsvetkov et al. (2008) \\
                                   & M1 III  & M2 III      & Derived\tablefootmark{a}  \\
    $\textit{$m_{v}$}$ [mag]  & 6.4    & 6.4  & Hipparcos (ESA 1997)  \\
    $\textit{$M_{v}$}$ [mag]  &  -- 1.24  & -- 1.52  & Famaey et al. (2005)      \\
    $\textit{B-V}$ [mag]      & 1.68             & 1.66             & Hipparcos (ESA 1997) \\
                              & 1.70 $\pm$ 0.02  & 1.68 $\pm$ 0.01  & van Leeuwen (2007)   \\
    $\textit{V-K}$ [mag]      & 4.14      & 4.39    & Cutri et al. (2003) \\
    age [Gyr]                 & 2.0 $\pm$ 1.3      & 4.5 $\pm$ 2.8  & Derived\tablefootmark{b}  \\
    Distance [pc]             & 320.2 $\pm$ 61.1  & 290.2 $\pm$ 45.5 & Famaey et al. (2005) \\
    RV [km s$^{-1}$]          & 4.79 $\pm$ 0.06 &  -- 36.89 $\pm$ 0.21 & Famaey et al. (2005)     \\
    Parallax [mas]            & 2.48 $\pm$ 0.38   & 3.08 $\pm$ 0.43  & van Leeuwen (2007) \\
    $T_{\mathrm{eff}}$ [K]    & 3950                   & 5080                  & Wright et al. (2003) \\
                              & 4643 $^{+598}_{-501}$  & 4971 $^{+741}_{-643}$ & Ammons et al. (2006) \\
                              & 4035 $\pm$ 65          & 3935 $\pm$ 110        & This work  \\
    $\mathrm{[Fe/H]}$         & -- 0.09 $\pm$ 0.16     & -- 0.25 $\pm$ 0.25    & This work  \\
    log $\it g$               &     1.6 $\pm$ 0.3      & 1.3 $\pm$ 0.5         & This work  \\
                              &     1.4 $\pm$ 0.2      & 1.1 $\pm$ 0.2         & Derived\tablefootmark{b}  \\
    $\textit{$R_{\star}$}$ [$R_{\odot}$] &  51.1 $\pm$ 8.3     & 49.7 $\pm$ 9.5  & Derived\tablefootmark{b} \\
    $\textit{$M_{\star}$}$ [$M_{\odot}$] & 1.6 $\pm$ 0.4     & 1.2 $\pm$ 0.3   & Derived\tablefootmark{b} \\
    $v_{\mathrm{rot}}$ sin $i$ [km s$^{-1}$]   &   3.6  &   3.0  &  This work  \\
    $P_{\mathrm{rot}}$ / sin $i$ [days]    & 660.0 $\pm$ 116.6 & 832.4 $\pm$ 156.7 & Derived\tablefootmark{a} \\
    $v_{\mathrm{micro}}$ [km s$^{-1}$]         &    1.7 $\pm$ 0.3   &  1.6 $\pm$ 0.3    &  This work  \\

\hline

\end{tabular}
\end{center}
\tablefoottext{a}{See text}.
\tablefoottext{b}{Derived using the online tool (http://stevoapd.inaf.it/cgi-bin/param}).
\end{table*}
For the past four years, we have conducted precise RV measurements of 40 K dwarfs. Here, we present two stars with long-period and low-amplitude RV variations. In Sect. 2, we describe the observations and data reduction. In Sect. 3, the stellar characteristics of the host stars are reinterpreted, including chromospheric activities and Hipparcos photometric variations. The RV measurements and orbital solutions will be presented in Sect. 4. Finally, in Sect. 5, we discuss our results from our study.
%


\section{Observations and reduction}

Since 2003, we have conducted a precise RV survey for $\sim$ 300 stars, including 55 K giants (Han et al. 2010), $\sim$ 200 G giants (Omiya et al. 2008), 10 M giants, and 40 K dwarfs.
The observations were carried out as part of the study on the 40 K dwarfs. We used a fiber-fed high-resolution ($\emph{R}$ = 45 000) Bohyunsan Observatory Echelle Spectrograph (BOES; Kim et al. 2007). The BOES spectra covered a wavelength region from 3500 to 10 500\,${\AA}$. In order to provide precise RV measurements of the K dwarfs, an iodine absorption (I$_{2}$) cell was used with a wavelength region of 4900 -- 6000 {\AA}. Each estimated S/N ratio at an I$_{2}$ region was about 200 with a typical exposure time ranging from 10 to 20 minutes. The long-term stability of the BOES was demonstrated by observing the RV standard star $\tau$ Ceti, which was constant with an rms scatter of 6.8 m s$^{-1}$ over the time span of our observations (Lee et al. 2011, 2012).

We obtained 24 and 32 spectra each for HD 208527 and HD 220074, respectively, from September 2008 to June 2012. The basic reduction of spectra was performed using the IRAF (Tody 1986) software package and DECH (Galazutdinov 1992) code. The precise RV measurements related to the I$_2$ analyses were undertaken using the RVI2CELL (Han et al. 2007), which is based on a method by Butler et al. (1996) and Valenti et al. (1995). The RV measurements for HD 208527 and HD 220074 are listed in Tables~\ref{tab2} and ~\ref{tab3} and shown in Figure~\ref{orbit1} and \ref{orbit2}.

%
   \begin{figure}
   \centering
   \includegraphics[width=8cm]{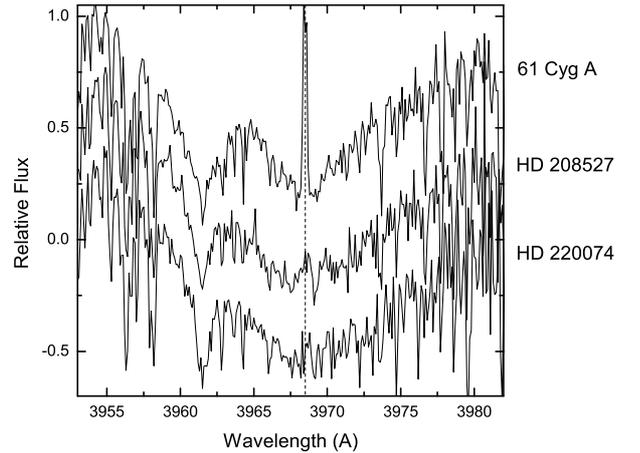}
      \caption{The Ca II H line cores for samples. The vertical dotted line indicates the center of the Ca II H regions. There are slight central emissions that exist in the center of the Ca II H core feature for HD 208527 and HD 220074, which are ambiguous to confirm the samples to be active. 61 Cyg A is shown for comparison.
        }
        \label{Ca1}
   \end{figure}
%

%
   \begin{figure*}
   \centering
   \includegraphics[width=12cm]{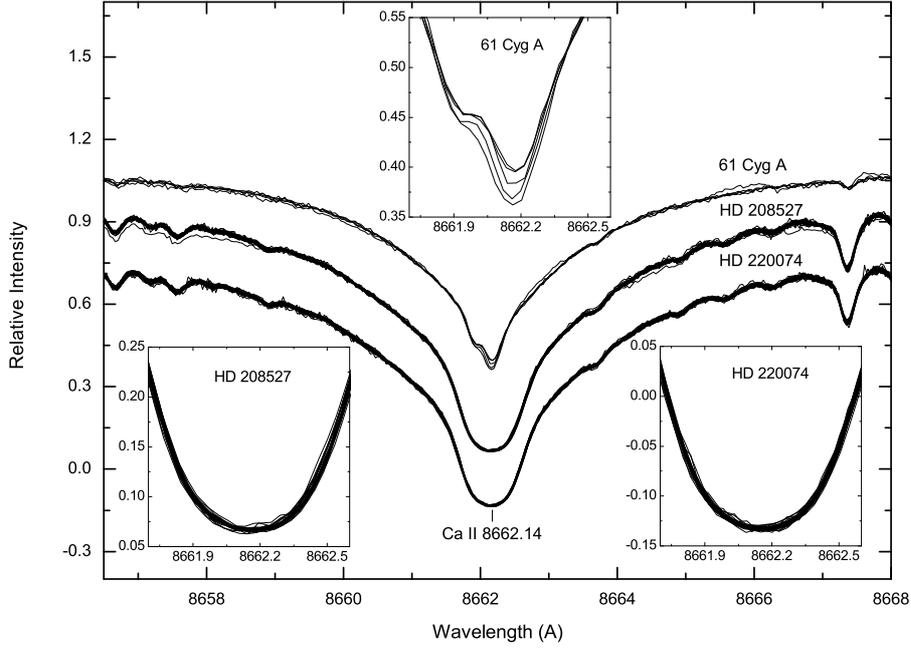}
      \caption{Line variations in the Ca II 8662 ${\AA}$ spectral region. While the active star 61 Cyg A shows the variations of EW, there are invisible variations for HD 208527 and HD 220074 at the Ca II 8662 ${\AA}$ core feature. They show typical line profiles of an inactive star.
        }
        \label{Ca2}
   \end{figure*}
%
%


\section{Stellar characteristics}

\subsection{Fundamental parameters}

The atmospheric parameters were determined based on the TGVIT (Takeda et al. 2005) code. To estimate the parameters, we used 169 (HD 208527) and 145 (HD 220074) equivalent width (EW) measurements of Fe I and Fe II lines. They resulted in $T_{\mathrm{eff}}$ = 4035 $\pm$ 65 K, log $\it g$ = 1.6 $\pm$ 0.3, $v_{\mathrm{micro}}$ = 1.7 $\pm$ 0.3 km s$^{-1}$ and $\mathrm{[Fe/H]}$ = -- 0.09 $\pm$ 0.16 for HD 208527, and $T_{\mathrm{eff}}$ = 3935 $\pm$ 110 K, log $\it g$ = 1.3 $\pm$ 0.5, $v_{\mathrm{micro}}$ = 1.6 $\pm$ 0.3 km s$^{-1}$ and $\mathrm{[Fe/H]}$ = -- 0.25 $\pm$ 0.25 for HD 220074. The projected rotational velocities were derived, $v_{\mathrm{rot}}$ sin $i$ = 3.6 and 3.0 km s$^{-1}$, from the widths of the spectral lines regarding the line-broadening functions. Based on the rotational velocities and the stellar radius of 51.1 $\pm$ 8.3 and 49.7 $\pm$ 9.5 $R_{\odot}$, we derived the range of the upper limit of the rotational period of 660.0  $\pm$ 116.6 (HD 208527) and 832.4 $\pm$ 156.7 (HD 220074) days.

Stellar masses were estimated from the theoretical stellar isochrones by using their position in the color--magnitude diagram based on the ones by Bertelli et al. (1994) and Girardi et al. (2000). We also adopted a version of the Bayesian estimation method (J{\o}rgensen \& Lindegren 2005; da Silva et al. 2006) by using the determined values for $T_{\mathrm{eff}}$, $\mathrm{[Fe/H]}$, $M_{v}$, and parallax. With our estimated parameters, we yielded a stellar mass of 1.6 $\pm$ 0.4 (HD 208527) and 1.2 $\pm$ 0.3 $M_{\odot}$ (HD 220074).
The basic stellar parameters are summarized in Table~\ref{tab1}.

HD 208527 ( = HR 8372 = HIP 108296) has been known to be a MS star with a spectral type of K5 V (ESA 1997) and an apparent magnitude of 6.4. The absolute magnitude of -- 1.24, however, indicates that the star could  be located at a giant stage in the H-R diagram. The values of the surface gravity and stellar radius also show that it may be a giant star. The surface gravity of the star is consistent with a giant but inconsistent with a dwarf. Stars that have the same spectral class but a different luminosity class have quite different optical versus infrared colours. Bessel \& Brett (1988) provide the intrinsic \emph{V-K} colours for dwarfs and giants. The \emph{V-K} colour of 4.14 is close to an M1 giant (\emph{V-K} = 4.05) rather than a K5 giant (\emph{V-K} = 3.60) according to Bessel \& Brett (1988). Thus, HD 208527 should be classified as an M giant rather than as a K dwarf star.

The classification of HD 220074 ( = HR 8881 = HIP 115218) has caused controversy. Whereas the result of the Hipparcos photometric observation estimated that the star remains in a MS stage with a spectral type of K1 V, recent reports showed a spectral type of M2 III (Kidger et al. 2003; Tsvetkov et al. 2008). The colour index of HD 220074 was estimated to be approximately 1.7 by Hipparcos data, which was reduced afresh by Leeuwen (2007) and shows the same result. It means that the star should be a spectral type of a late K or early M.
However, the effective temperatures have led to confusion over the spectral classification of HD 220074. Studies have shown it to be $\sim$ 5000 K (Wright et al. 2003; Ammons et al. 2006), which indicates the spectral type of an early K and it agrees with the Hipparcos classification of a K1-type. On the contrary, our estimation of an effective temperature ($T_{\mathrm{eff}}$ = 3935 K) shows that the spectral type is close to a late K or an early M. HD 220074 also could be classified as a giant star according to the value of the absolute magnitude $\textit{$M_{v}$}$ = -- 1.52, as well as a low surface gravity and a large stellar radius compared to that of the dwarf star. The \emph{V-K} colour of 4.39 is similar to an M2 giant (\emph{V-K} = 4.30) in Bessel \& Brett (1988), which also indicates that the star is likely to be a giant. As a result, HD 220074 should be an M2 giant star (Kidger et al. 2003; Tsvetkov et al. 2008).

\subsection{Chromospheric activities}



The EW variations of the Ca II H \& K lines are frequently used as chromospheric activity indicators because they are sensitive to stellar activity and the variations could have a decisive effect on the RV variations. The emissions in the Ca II H \& K core are formed in the chromosphere and show a typical central reversal in the presence of chromospheric activity (Pasquini et al. 1988; Saar \& Donahue 1997). Figure~\ref{Ca1} shows individual features in the Ca II H line region. Due to a low S/N of the blue-most region, the spectrum is not clear enough to resolve the emission feature in the Ca II H line core. It may be the result of some invisible emission features or scatters in the both samples. Even if there really are slight central emissions compared to 61 Cyg A (K5 V) showing atmospheric activity, which is one of the samples of 40 K dwarfs, the features are too weak to confirm the existence of core emission features categorically.

Chromospheric activity also has been measured by using the Ca II triplet lines (Larson et al. 1993; Hatzes et al. 2003; Han et al. 2010). Among the triplet lines (8498, 8542, and 8662 ${\AA}$) , Ca II 8662 ${\AA}$ is uncontaminated by atmospheric lines near the core. Changes in the core flux of the Ca II 8662 ${\AA}$ reflect variations in stellar chromospheric activity and qualitatively related to the variations in the Ca II H \& K flux (Larson et al. 1993). The line profiles in the Ca II 8662 ${\AA}$ spectral region are shown in Figure~\ref{Ca2}. The small boxes are magnified regions of 1 ${\AA}$ centered on the Ca II 8662 ${\AA}$ to show the EW variations near the core. The comparison of 61 Cyg A, known as the active star, was obtained during the same season at the epoch (JD-2454726.198, 2454726.220, 2454752.124, 2454833.894, and 2455018.172). The spectra of 61 Cyg A clearly show variations of the central core, which is very similar to the result in Larson et al. (1993). On the other hand, the line features for HD 208527 and HD 220074 display no such variations.

%
\subsection{Photometric variations}

\begin{figure}
   \centering
   \includegraphics[width=8cm]{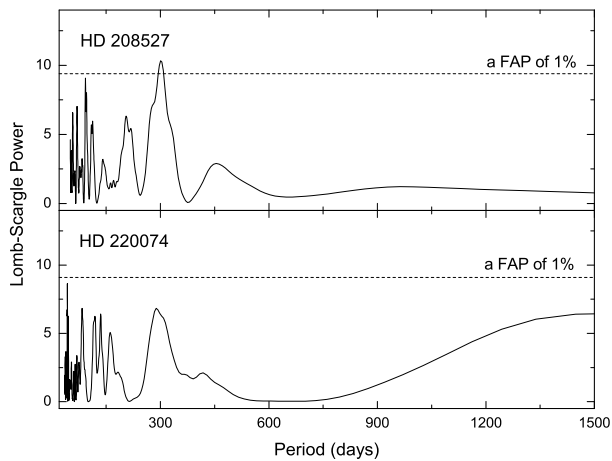}
      \caption{The Hipparcos photometric variations for HD 208527 (\emph{top panel}) and  HD 220074 (\emph{bottom panel}), respectively. The horizontal line indicates a FAP threshold of 1\%.
        }
        \label{Hip}
   \end{figure}

We analyzed the Hipparcos photometry data to search for possible brightness variations, which may be caused by the rotational modulation of cool stellar spots. The available photometry database originates from 88 Hipparcos measurements for HD 208527 (from November 1989 to December 1992) and 121 Hipparcos measurements for HD 220074 (from January 1990 to December 1992). They maintained a photometric stability down to rms scatters of 0.010 (HD 208527) and 0.012 (HD 220074) magnitude, which correspond to 0.15\% and 0.18\% variations over the time span of the observations. Figure~\ref{Hip} shows the Lomb-Scargle periodograms (Lomb 1976; Scargle 1982) of the Hipparcos data for both stars. A rather large peak appears at about 301 days with a false alarm probability (FAP) of 0.2 $\pm$ 0.1\% (HD 208527). A FAP is a metric to express the significance of a period by a bootstrap randomization technique (Press et al. 1992; K\"{u}rster et al. 1999). But, there are no additional significant peaks at other periods in both periodograms.





%
%
%


\section{Radial velocity variations and origin}

\subsection{HD 208527}

%
\begin{table}
\begin{center}
\caption{RV measurements for HD 208527 from September 2008 to June 2012 using the BOES.}
\label{tab2}
\begin{tabular}{cccccc}
\hline\hline

 JD         & $\Delta$RV  & $\pm \sigma$ &        JD & $\Delta$RV  & $\pm \sigma$  \\
 -2 450 000 & m\,s$^{-1}$ &  m\,s$^{-1}$ & -2 450 000  & m\,s$^{-1}$ &  m\,s$^{-1}$  \\
\hline

4736.119458 &   -91.0 &   16.5 & 5454.126840  &    87.2  &     9.3   \\
4752.152971 &   -84.2 &   11.3 & 5697.263528  &  -148.0  &    20.2   \\
4833.916842 &  -176.6 &    8.0 & 5698.288703  &   -98.7  &    10.8   \\
5018.194241 &   -82.3 &   11.0 & 5711.267459  &  -125.7  &    12.8   \\
5083.979529 &    -3.6 &    8.6 & 5820.129124  &   -80.6  &     9.4   \\
5131.919794 &    -5.7 &   10.9 & 5842.035628  &  -208.3  &    11.1   \\
5169.956708 &    69.6 &   11.1 & 5911.006597  &  -138.1  &    10.7   \\
5183.885600 &    53.1 &   10.2 & 6024.330791  &    51.7  &    12.1   \\
5224.889144 &   129.5 &   19.0 & 6026.309597  &    38.4  &    15.3   \\
5311.277212 &    79.7 &   10.4 & 6069.265534  &   134.2  &    11.5   \\
5321.264398 &   120.4 &   11.1 & 6086.197407  &   202.6  &    30.0   \\
5356.307074 &   183.2 &    9.2 & 6088.318959  &    94.0  &    15.1   \\
\hline

\end{tabular}
\end{center}
\end {table}
%

%
   \begin{figure}
   \centering
   \includegraphics[width=8cm]{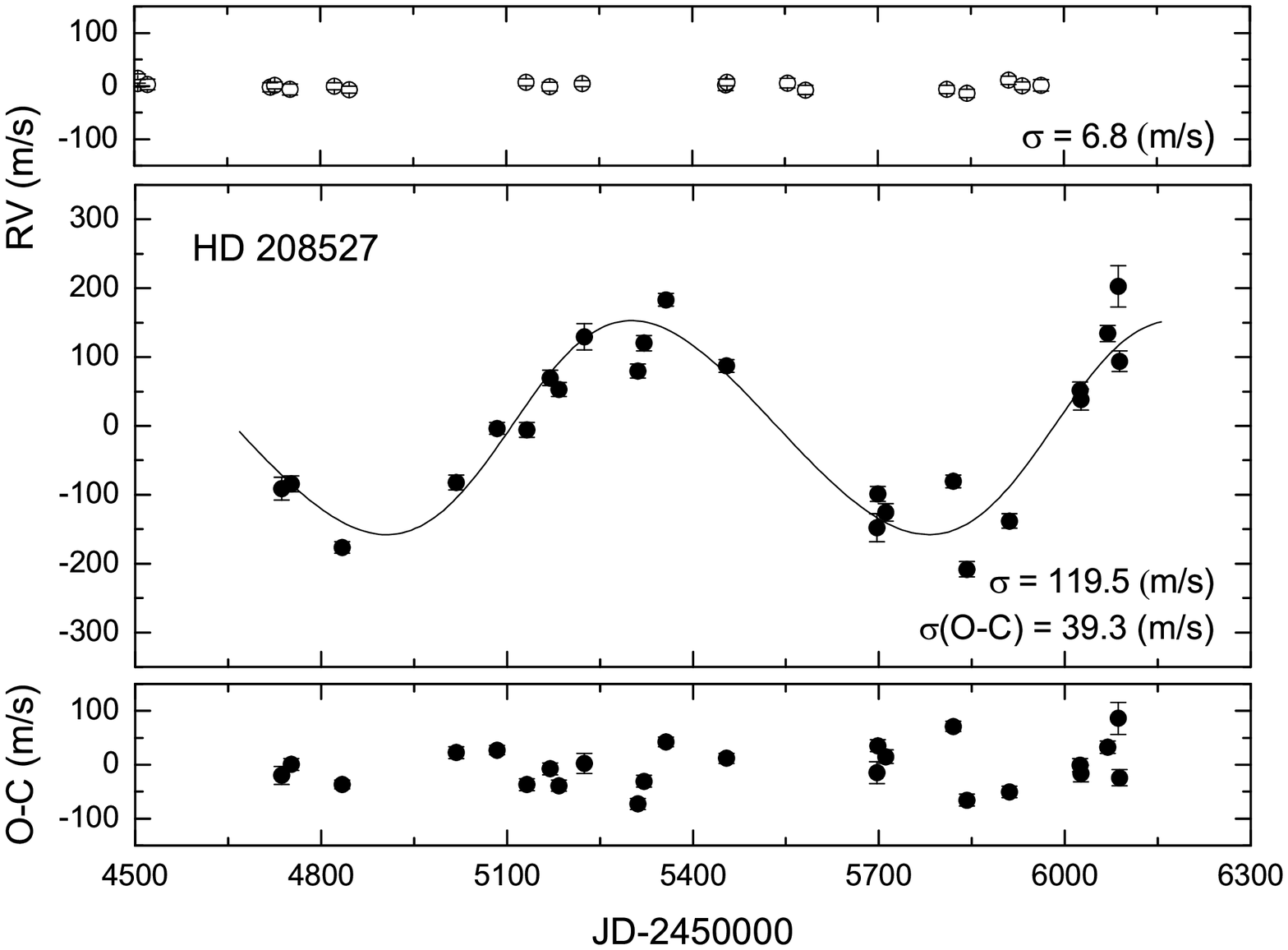}
      \caption{Variations of the RV standard $\tau$ Ceti (\emph{top panel}), RV curve (\emph{middle panel}), and rms scatter of the residual (\emph{bottom panel}) for HD 208527 from September 2008 to June 2012. The solid line is the orbital solution with a period of 875 days and an eccentricity of 0.08.
              }
         \label{orbit1}
   \end{figure}
%

%
   \begin{figure}
   \centering
   \includegraphics[width=8cm]{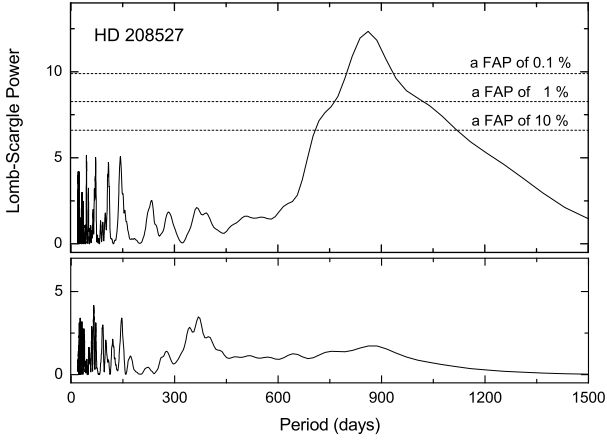}
      \caption{A Lomb-Scargle periodogram of the RV measurements for HD 208527.
      The periodogram shows a significant power at a period of 875.5 days (\emph{top panel}) and after subtracting the main frequency variations (\emph{bottom panel}). The horizontal lines indicate FAP thresholds of 0.1\%, 1\%, and 10\% (\emph{top} to \emph{bottom}).
      }
         \label{power1}
   \end{figure}

Period analysis was applied to appraise the significance of periodic trends or variabilities of the time series. A Lomb-Scargle periodogram, which is a useful tool to investigate long-period variations for unequally spaced data, was used to calculate the RV time series. The time series of our RV measurements for HD 208527 are listed in Table~\ref{tab2} and shown in Figure~\ref{orbit1} (middle panel). Figure~\ref{power1} shows a significant power for HD 208527 at a period of 875.5 days (top panel) with a FAP of less than $10^{-5}$ and the residual after subtracting the superior period (bottom panel). There is a highly significant power at a period of 875.5 days and no significant peak exists in the residual.

\begin{figure}
   \centering
   \includegraphics[width=8cm]{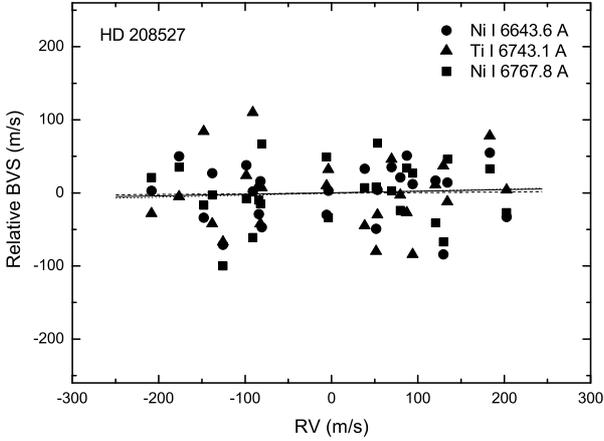}
      \caption{The examinations of the origin of the BVS variations for HD 208527.
      It shows the relations between the RV and BVS variations for three spectral lines (Ni I 6643.6, Ti I 6743.1, and Ni I 6767.8 {\AA}). The solid, dashed, and dotted lines mark the slopes of the three spectral lines, respectively.
        }
        \label{BVS1}
   \end{figure}

HD 208527 is reclassified as a giant star in this work, rather than a dwarf. Thus, HD 208527 may have surface inhomogeneities which are not uncommon to giant stars. Stellar rotational modulations of surface features can create variable asymmetries in the spectral line profiles (Queloz et al. 2001), and in particular, the variations in the shapes of the spectral line may cause the RV variations. The different RV measurements between the high and low flux levels of the line profile are defined as a bisector velocity span (BVS). In order to calculate the BVS, we selected three strong lines with a high flux level and unblended spectral features, which are Ni I 6643.6, Ti I 6743.1, and Ni I 6767.8 {\AA}. The lines are located beyond the I$_{2}$ absorption region and show a high flux level throughout the whole wavelength region. We estimated each BVS of the profile between two different flux levels with central depth levels of 0.8 and 0.4 as the span points, that avoided the spectral core and wing where the difference of the bisector measurements are large. The BVS variations of HD 208527 as a function of RV are shown in Figure~\ref{BVS1}. The slopes are 0.02 (Ni I 6643.6), 0.01 (Ti I 6743.1), and 0.03 (Ni I 6767.8 {\AA}), respectively; There are no obvious correlations between the RV and BVS measurements.

HD 208527 reveals a periodic signal in RV variations, and the best fit Keplerian orbit is found with a $P$ = 875.5 $\pm$ 5.8 days, a $K$ = 155.4 $\pm$ 3.2 m s$^{-1}$, and an $e$ = 0.08 $\pm$ 0.04.
Figure~\ref{orbit1} shows the RV curve as a function of time for HD 208527 and the residual after extracting the main frequency. Assuming a stellar mass of 1.6 $\pm$ 0.4 $M_{\odot}$, we derived a minimum mass of a planetary companion $m$ sin $i$ = 9.9 $\pm$ 1.7 $\it M_\mathrm{Jup}$ at a distance $a$ = 2.1 AU from HD 208527.

%
%
%

%
%
%

%
\subsection{HD 220074}

\begin{figure}
   \centering
   \includegraphics[width=8cm]{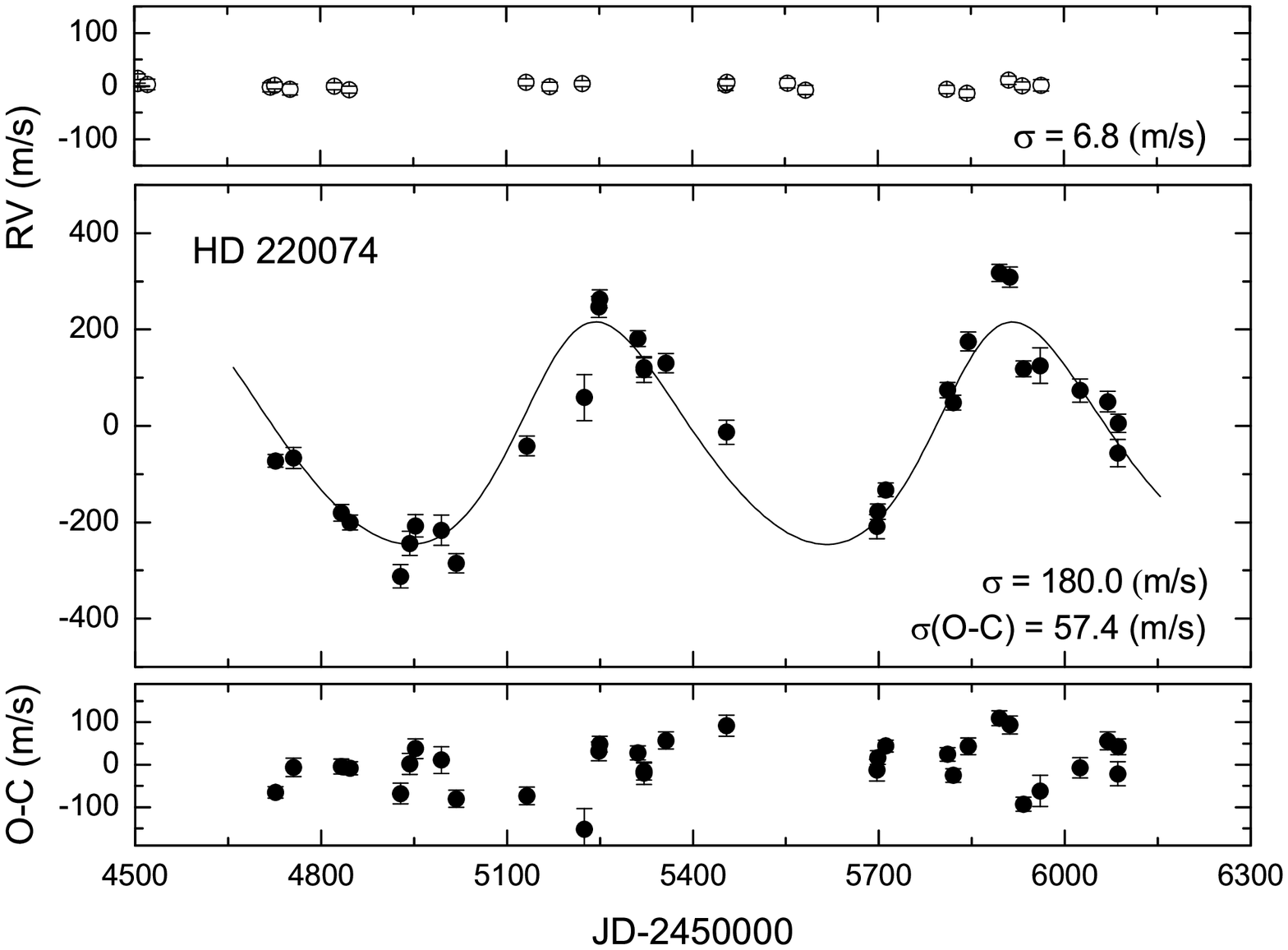}
      \caption{Variations of the RV standard $\tau$ Ceti (\emph{top panel}), RV curve (\emph{middle panel}), and rms scatter of the residual (\emph{bottom panel}) for HD 220074 from September 2008 to June 2012. The solid line is the orbital solution with a period of 672 days and an eccentricity of 0.14.
              }
         \label{orbit2}
   \end{figure}

\begin{figure}
   \centering
   \includegraphics[width=8cm]{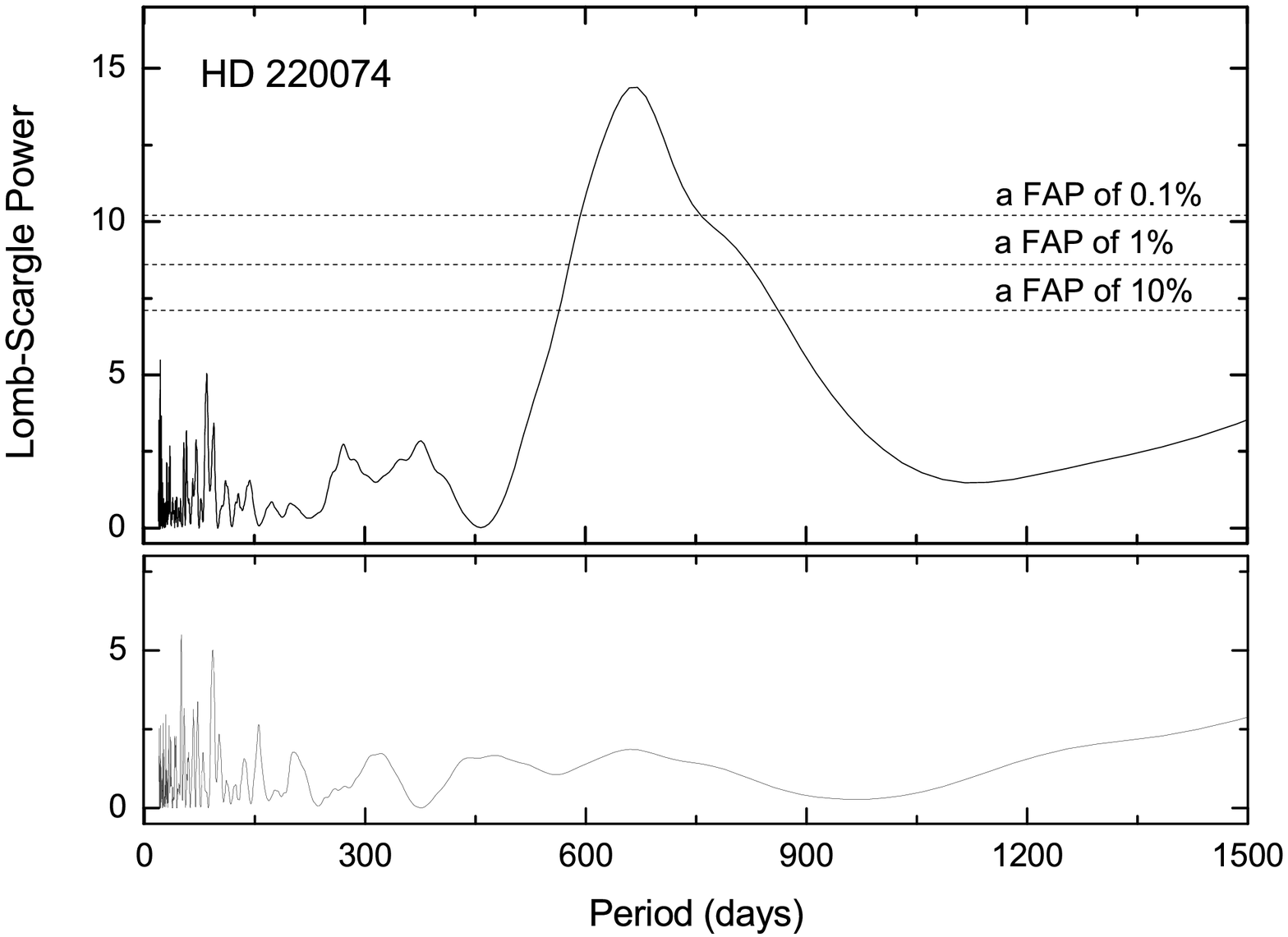}
      \caption{A Lomb-Scargle periodogram of the RV measurements for HD 220074.
      The periodogram shows a significant power at a period of 672.1 days (\emph{top panel}) and after subtracting the main frequency variations (\emph{bottom panel}). The horizontal lines indicate FAP thresholds of 0.1\%, 1\%, and 10\% (\emph{top} to \emph{bottom}), respectively.
      }
         \label{power2}
   \end{figure}

The time series of our RV measurements for HD 220074 are listed in Table~\ref{tab3} and shown in Figure~\ref{orbit2}. A Lomb-Scargle periodogram of the data exhibits a dominant peak at a period of 672.1 days in Figure~\ref{power2} (top panel) with a FAP of less than $10^{-5}$. Figure~\ref{power2} (bottom panel) shows a Lomb-Scargle periodogram of the residual after subtracting the superior period and no significant peak exists.

\begin{figure}
   \centering
   \includegraphics[width=8cm]{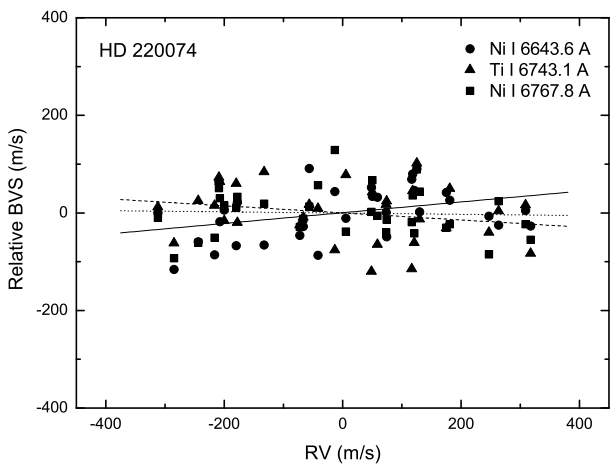}
      \caption{The examinations of the origin of the BVS variations for HD 220074.
      It shows the relations between the RV and BVS variations for three spectral lines (Ni I 6643.6, Ti I 6743.1, and Ni I 6767.8 {\AA}). The solid, dashed, and dotted lines mark the slopes of the three spectral lines, respectively.
        }
        \label{BVS2}
   \end{figure}

The BVS was measured by using the same spectral features and method as that for HD 208527. The BVS variations of HD 220074 as a function of RV are shown in Figure~\ref{BVS2}. The slopes are 0.10 (Ni I 6643.6), -- 0.07 (Ti I 6743.1), and -- 0.01 (Ni I 6767.8 {\AA}), respectively. The correlation between the BVS and RV is not visible. However, we want to point out that this can also be due to too small BVS variations, as expected from Saar \& Donahue (1997) or Hatzes (2002).

The RV variations of HD 220074 can be well-reproduced by Keplerian orbit with a $P$ = 672.1 $\pm$ 3.7 days, a $K$ = 230.8 $\pm$ 5.0 m s$^{-1}$, and an $e$ = 0.14 $\pm$ 0.05. The RV variations and residuals for HD 220074 are shown in Figure~\ref{orbit2}. Assuming a stellar mass of 1.2 $\pm$ 0.3 $M_{\odot}$,  we derived a minimum mass of a planetary companion $m$ sin $i$ = 11.1 $\pm$ 1.8 $\it M_\mathrm{Jup}$ at a distance $a$ = 1.6 AU from HD 220074. All the orbital elements are listed in Table~\ref{tab4}.

%
\begin{table}
\begin{center}
\caption{RV measurements for HD 220074 from September 2008 to June 2012 using the BOES.}
\label{tab3}
\begin{tabular}{cccccc}
\hline\hline

 JD         & $\Delta$RV  & $\pm \sigma$ &        JD & $\Delta$RV  & $\pm \sigma$  \\
 -2 450 000 & m\,s$^{-1}$ &  m\,s$^{-1}$ & -2 450 000  & m\,s$^{-1}$ &  m\,s$^{-1}$  \\
\hline

4726.297467  &    -72.5 &   13.5 &  5356.296709 &   130.1 &   20.2  \\
4755.202115  &    -66.7 &   21.8 &  5454.046174 &   -13.1 &   25.1  \\
4832.974465  &   -180.0 &   17.2 &  5697.274048 &  -208.9 &   25.3  \\
4846.945322  &   -200.0 &   15.5 &  5698.299255 &  -178.0 &   15.7  \\
4928.351174  &   -312.3 &   24.2 &  5711.297204 &  -132.6 &   14.0  \\
4943.312332  &   -243.9 &   24.9 &  5811.077022 &    74.4 &   16.2  \\
4952.326188  &   -207.5 &   23.1 &  5820.155239 &    48.6 &   15.6  \\
4994.232534  &   -216.5 &   31.3 &  5844.277787 &   175.3 &   19.4  \\
5018.233079  &   -285.0 &   20.2 &  5894.184461 &   317.6 &   17.6  \\
5131.961073  &    -41.7 &   20.5 &  5911.065978 &   309.0 &   21.0  \\
5224.901848  &     58.9 &   48.0 &  5933.008989 &   118.2 &   16.6  \\
5247.906226  &    247.1 &   21.9 &  5959.959457 &   125.2 &   37.0  \\
5249.914037  &    263.5 &   18.8 &  6024.351659 &    73.5 &   23.9  \\
5311.263220  &    181.1 &   16.5 &  6069.282782 &    50.4 &   21.3  \\
5321.251055  &    116.9 &   26.7 &  6085.211924 &   -56.5 &   28.5  \\
5321.276889  &    121.1 &   20.1 &  6086.226927 &     5.5 &   18.8  \\

\hline

\end{tabular}
\end{center}
\end {table}
%

%
%

%
%
%

%
%
%

%
%
%


\section{Discussion}

In order to search for and study the origin of planetary companions, the precise RV measurements around 40 K dwarfs were obtained with BOES for four years. We find that HD 208527 and HD 220074 show evidences of low-amplitude and long-periodic RV variations. The spectral and luminosity classes for the samples are reclassified with the result of previous studies and our new estimation of the stellar parameters. HD 208527, known as a K dwarf, is changed to an M giant and HD 220074, known as a K dwarf or an M giant, is confirmed to be an M giant. Thus, the low-amplitude RV variations may have been caused by the surface phenomena as well as by orbiting planetary companions.

We found RV periods of 875 and 672 days for HD 208527 and HD 220074, respectively. Long-term RV variations, such as in hundreds of days, in evolved stars may have been caused by three kinds of phenomena:
stellar pulsations, rotational modulations by inhomogeneous surface features, or planetary companions.
Of these, RV variations due to pulsations are unlikely because the period of the fundamental radial mode pulsation is several days and the mechanism to produce very long and low-amplitude pulsation modes is impractical.

The lack of activity in the stars argue in favor of the planet hypothesis. The Ca II H \& K line and Ca II 8662 ${\AA}$ were used to monitor chromospheric activities, which moreover show strong correlations from the results of early K stars (Thatcher \& Robinson 1993). Because of the low S/N in the Ca II H \& K region of the BOES spectra, the variations in the Ca II 8662 ${\AA}$ line profile were examined instead: they do not show any notable variations. 
The Hipparcos photometric variations and the BVS also showed no correlations with the RV measurements. Even though there exists a moderate power at a frequency of 301 days, as compared to 875 days seen in RV variations, in Hipparcos photometric data for HD 208527, this does not go against the general trend. All these null results from the analyses of the chromospheric activities or surface inhomogeneities argues against the rotational modulation of surface activities as the origin of observed RV variations.

However, the existence of planetary companions can explain the observed RV variations. Assuming the stellar masses of 1.6 $\pm$ 0.4 and  1.2 $\pm$ 0.3 $M_{\odot}$ for HD 208527 and HD 220074, the minimum masses of planetary companions were estimated to be 9.9 $\pm$ 1.7 and 11.1 $\pm$ 1.8 $\it M_\mathrm{Jup}$ with an orbital semi-major axis of 2.1 and 1.6 AU, respectively. If confirmed, HD 208527 and HD 220074 become the first candidates harboring planetary companions around M giant stars.

After removing the RV signal of the planets, the remaining RV variations are 39.3 m s$^{-1}$ (HD~208527) and 57.4 m s$^{-1}$ (HD~220074), respectively. The variations are larger than the expected for stellar activity and the errors of the measurements. Additional frequencies can be considered and duly weighed because the values are significantly larger than the RV precision for the RV standard star $\tau$ Ceti (6.8 m s$^{-1}$), as well as larger than the typical internal error of individual RV accuracies of $\sim$ 13 m s$^{-1}$ (HD 208527) and $\sim$ 22 m s$^{-1}$ (HD 220074). A periodogram of the RV residual variations, however, does not show any additional periodic signal as shown in Figure~\ref{power1} (bottom panel) and Figure~\ref{power2} (bottom panel). This means that the residual RV variations do not indicate the presence of another planet. Hekker et al. (2006) have shown that the rms of RV residuals of K giants has a median value of 20 m s$^{-1}$ and this tends to increase toward later spectral types. Hatzes et al. (2005) also have shown that the high rms of residuals ($\sim$ 51 m s$^{-1}$) are found in the K1 giant HD 13189 and this shows a significant short-term RV variability on time scales of days that is most likely due to stellar oscillations. That behavior is typical for K giant stars, whether periodic or not (Setiawan et al. 2003; Hatzes et al. 2005; D{\"o}linger et al. 2007; Lee et al. 2008; de Medeiros et al. 2009; Han et al. 2010).


%
\begin{table}
\begin{center}
\caption{Orbital parameters for HD 208527 b and HD 220074 b.}
\label{tab4}
\begin{tabular}{lcc}
\hline
\hline
    Parameter                            & HD 208527               & HD 220074             \\

\hline
    Period [days]                        & 875.5  $\pm$ 5.8       & 672.1  $\pm$ 3.7     \\
    $\it T$$_{\mathrm{periastron}}$ [JD] & 2450745.3 $\pm$ 45.7   & 2451158.2 $\pm$ 31.7 \\
    $\it{K}$ [m s$^{-1}$]                & 155.4  $\pm$ 3.2       & 230.8  $\pm$ 5.0     \\
    $\it{e}$                             & 0.08   $\pm$ 0.04      & 0.14   $\pm$ 0.05    \\
    $\omega$ [deg]                       & 278.0   $\pm$ 19.5     & 323.0   $\pm$ 8.5    \\
    $f(m)$ [$\it M_{\odot}$]             & (3.375) $\times$ 10$^{-7}$    & (8.314) $\times$ 10$^{-7}$  \\
    $a$ sin $i$ [AU]                     & (1.247) $\times$ 10$^{-2}$    & (1.412) $\times$ 10$^{-2}$  \\
    $\sigma$ (O-C) [m s$^{-1}$]          & 39.3                   & 57.4                 \\
\hline
    $m$ sin $i$ [$\it M_\mathrm{Jup}$]   & 9.9 $\pm$ 1.7    & 11.1 $\pm$ 1.8    \\
    $\it{a}$ [AU]                        & 2.1 $\pm$ 0.2    &  1.6 $\pm$ 0.1    \\
\hline

\end{tabular}
\end{center}
\end{table}
%


\begin{acknowledgements}
      BCL acknowledges partial support by the KASI (Korea Astronomy and Space Science Institute) grant 2012-1-410-03. Support for MGP was provided by the National Research Foundation of Korea to the Center for Galaxy Evolution Research.
      This research made use of the SIMBAD database, operated at CDS, Strasbourg, France.

\end{acknowledgements}
%



\begin{thebibliography}{}

\bibitem[Ammons et al.(2006)]{2006ApJ...638.1004A} Ammons, S.~M., Robinson, S.~E., Strader, J., et al.\ 2006, \apj, 638, 1004

\bibitem[Bertelli et al.(1994)]{1994A&AS..106..275B} Bertelli, G., Bressan, A., Chiosi, C., et al.\ 1994, \aaps, 106, 275
\bibitem[Bessell \& Brett(1988)]{1988PASP..100.1134B} Bessell, M.~S., \& Brett, J.~M.\ 1988, \pasp, 100, 1134

\bibitem[Bonfils et al.(2011)]{2011arXiv1111.5019B} Bonfils, X., Delfosse, X., Udry, S., et al.\ 2011, arXiv:1111.5019

\bibitem[Butler et al.(1996)]{1996PASP..108..500B} Butler, R.~P., Marcy, G.~W., Williams, E., et al.\ 1996, \pasp, 108, 500

\bibitem[Cutri et al.(2003)]{2003yCat.2246....0C} Cutri, R.~M., Skrutskie, M.~F., van Dyk, S., et al.\ 2003, VizieR Online Data Catalog, 2246, 0

\bibitem[da Silva et al.(2006)]{2006A&A...458..609D} da Silva, L., Girardi, L., Pasquini, L., et al.\ 2006, \aap, 458, 609

\bibitem[de Medeiros et al.(2009)]{2009A&A...504..617D} de Medeiros, J.~R., Setiawan, J., Hatzes, A.~P., et al.\ 2009, \aap, 504, 617

\bibitem[D{\"o}llinger et al.(2007)]{2007A&A...472..649D} D{\"o}llinger, M.~P., Hatzes, A.~P., Pasquini, L., et al.\ 2007, \aap, 472, 649


\bibitem[Endl et al.(2003)]{2003AJ....126.3099E} Endl, M., Cochran, W.~D., Tull, R.~G., \& MacQueen, P.~J.\ 2003, \aj, 126, 3099

\bibitem[ESA(1997)]{1997yCat.1239....0E} ESA 1997, VizieR Online Data Catalog, 1239, 0

\bibitem[Famaey et al.(2005)]{2005A&A...430..165F} Famaey, B., Jorissen, A., Luri, X., et al.\ 2005, \aap, 430, 165

\bibitem[Freytag \& H{\"o}fner(2008)]{2008A&A...483..571F} Freytag, B., \& H{\"o}fner, S.\ 2008, \aap, 483, 571

\bibitem[Frink et al.(2002)]{2002ApJ...576..478F} Frink, S., Mitchell, D.~S., Quirrenbach, A., et al.\ 2002, \apj, 576, 478

\bibitem[Galazutdinov (1992)]{} Galazutdinov, G. A. 1992, Special Astrophysical Observatory Preprint 92 (Nizhnij Arkhyz: SAO)

\bibitem[Girardi et al.(2000)]{2000A&AS..141..371G} Girardi, L., Bressan, A., Bertelli, G., et al.\ 2000, \aaps, 141, 371


\bibitem[Gregory \& Fischer(2010)]{2010MNRAS.403..731G} Gregory, P.~C., \& Fischer, D.~A.\ 2010, \mnras, 403, 731

\bibitem[Han et al.(2007)]{2007PKAS...22...75H} Han, I., Kim, K.-M., Lee, B.-C., et al.\ 2007, PKAS, 22, 75

\bibitem[Han et al.(2010)]{2010A&A...509A..24H} Han, I., Lee, B.~C., Kim, K.~M., et al.\ 2010, \aap, 509, A24

\bibitem[Hatzes(2002)]{2002AN....323..392H} Hatzes, A.~P.\ 2002, Astronomische Nachrichten, 323, 392

\bibitem[Hatzes et al.(2003)]{2003ApJ...599.1383H} Hatzes, A.~P., Cochran, W.~D., Endl, M., et al.\ 2003, \apj, 599, 1383

\bibitem[Hatzes et al.(2005)]{2005A&A...437..743H} Hatzes, A.~P., Guenther, E.~W., Endl, M., et al.\ 2005, \aap, 437, 743

\bibitem[Hekker et al.(2006)]{2006A&A...454..943H} Hekker, S., Reffert, S., Quirrenbach, A., et al.\ 2006, \aap, 454, 943




\bibitem[J{\o}rgensen \& Lindegren(2005)]{2005A&A...436..127J} J{\o}rgensen, B.~R., \& Lindegren, L.\ 2005, \aap, 436, 127


\bibitem[Kidger \& Mart{\'{\i}}n-Luis(2003)]{2003AJ....125.3311K} Kidger, M.~R., \& Mart{\'{\i}}n-Luis, F.\ 2003, \aj, 125, 3311

\bibitem[Kim et al.(2007)]{2007PASP..119.1052K} Kim, K.-M., Han, I., Valyavin, G.~G., et al.\ 2007, \pasp, 119, 1052
\bibitem[Koen \& Laney(2000)]{2000MNRAS.311..636K} Koen, C., \& Laney, D.\ 2000, \mnras, 311, 636

\bibitem[K{\"u}rster et al.(1999)]{1999A&A...344L...5K} K{\"u}rster, M., Hatzes, A.~P., Cochran, W.~D., et al.\ 1999, \aap, 344, L5


\bibitem[Larson et al.(1993)]{1993PASP..105..332L} Larson, A.~M., Irwin, A.~W., Yang, S.~L.~S., et al.\ 1993, \pasp, 105, 332



\bibitem[Lee et al.(2008)]{2008AJ....135.2240L} Lee, B.-C., Mkrtichian, D.~E., Han, I., et al.\ 2008, \aj, 135, 2240

\bibitem[Lee et al.(2011)]{2011A&A...529A.134L} Lee, B.-C., Mkrtichian, D.~E., Han, I., et al.\ 2011, \aap, 529, A134

\bibitem[Lee et al.(2012)]{2012A&A...543A..37L} Lee, B.-C., Han, I., Park, M.-G., et al.\ 2012, \aap, 543, A37

\bibitem[Lomb(1976)]{1976Ap&SS..39..447L} Lomb, N.~R.\ 1976, \apss, 39, 447




\bibitem[Omiya et al.(2008)]{2008IAUS..249...53O} Omiya, M., Izumiura, H., Sato, B., et al.\ 2008, IAU Symposium, 249, 53

\bibitem[Pasquini et al.(1988)]{1988A&A...191..253P} Pasquini, L., Pallavicini, R., \& Pakull, M.\ 1988, \aap, 191, 253

\bibitem[Press et al.(1992)]{1992nrfa.book.....P} Press, W.~H., Teukolsky, S.~A., Vetterling, W.~T., et al.\ 1992, Cambridge: University Press, |c1992, 2nd ed.

\bibitem[Queloz et al.(2001)]{2001A&A...379..279Q} Queloz, D., Henry, G.~W., Sivan, J.~P., et al.\ 2001, \aap, 379, 279



\bibitem[Saar \& Donahue(1997)]{1997ApJ...485..319S} Saar, S.~H., \& Donahue, R.~A.\ 1997, \apj, 485, 319

\bibitem[Sato et al.(2003)]{2003ApJ...597L.157S} Sato, B., Ando, H., Kambe, E., et al.\ 2003, \apjl, 597, L157

\bibitem[Scargle(1982)]{1982ApJ...263..835S} Scargle, J.~D.\ 1982, \apj, 263, 835


\bibitem[Setiawan et al.(2003)]{2003A&A...398L..19S} Setiawan, J., Hatzes, A.~P., von der L{\"u}he, O., et al.\ 2003, \aap, 398, L19


\bibitem[Takeda et al.(2005)]{2005PASJ...57...27T} Takeda, Y., Ohkubo, M., Sato, B., et al.\ 2005, \pasj, 57, 27

\bibitem[Thatcher \& Robinson(1993)]{1993MNRAS.262....1T} Thatcher, J.~D., \& Robinson, R.~D.\ 1993, \mnras, 262, 1

\bibitem[Tody(1986)]{1986SPIE..627..733T} Tody, D.\ 1986, \procspie, 627, 733

\bibitem[Tsvetkov et al.(2008)]{2008AstL...34...17T} Tsvetkov, A.~S., Popov, A.~V., \& Smirnov, A.~A.\ 2008, Astronomy Letters, 34,~17

\bibitem[Valenti et al.(1995)]{1995PASP..107..966V} Valenti, J.~A., Butler, R.~P., \& Marcy, G.~W.\ 1995, \pasp, 107, 966

\bibitem[van Leeuwen(2007)]{2007A&A...474..653V} van Leeuwen, F.\ 2007, \aap, 474, 653

\bibitem[Wright et al.(2003)]{2003AJ....125..359W} Wright, C.~O., Egan, M.~P., Kraemer, K.~E., et al.\ 2003, \aj, 125, 359







\end{thebibliography}
\end{document}